\documentclass[10pt,conference]{IEEEtran}

\usepackage{graphicx} 
\usepackage{float}
\usepackage{url}
\usepackage[unicode]{hyperref}
\hypersetup{
   unicode=true,          
   plainpages=false,
   colorlinks=true,       
   citecolor=blue,        
}

\title{Popularity and Innovation in Maven Central}

\author{\IEEEauthorblockN{Nkiru Ede, Jens Dietrich, and Ulrich
Z\"ulicke}
\IEEEauthorblockA{\{nkiru.ede,jens.dietrich,uli.zuelicke\}@vuw.ac.nz\\
Victoria University of Wellington \\
Wellington, New Zealand}}

\begin{document}
\maketitle

\begin{abstract}  
Maven Central is a large popular repository of Java components
that has evolved over the last 20 years. The distribution of
dependencies indicates that the repository is dominated by a
relatively small number of components other components depend
on. The question is whether those elites are static, or change
over time, and how this relates to innovation in the Maven
ecosystem. We study those questions using several metrics. We
find that elites are dynamic, and that the rate of innovation
is slowing as the repository ages but remains healthy. 
\end{abstract}

\section{Introduction}
\label{section:introduction}

Software re-use has been revolutionized by the emergence of
software ecosystems (SECOs)~\cite{mens2023seco}. SECOs provide
an infrastructure to rapidly release, discover and use software
components. They are usually linked to build tools that include
dependency managers. Those tools can resolve symbolic references to
components in a SECO to physical references by downloading and
managing the respective components, often adding additional
functionality like security checks and conflict resolution. The
availability of standard formats used to declare dependencies
allows the study of SECOs as networks, where components are
modelled as vertices and dependencies between components as
edges~\cite{Valverde2002,Myers2003,Zheng2008,Cataldo2014}.

One such ecosystem is Maven. Maven repositories are used to
distribute binaries in the Java byte code format, produced by 
projects using Java, Kotlin and other languages that can be
compiled to Java byte code. There are numerous build tools and package
managers that can be used to interact (publish, query, etc.)
with Maven repositories,  \textit{maven} and \textit{gradle}
being the most popular ones. The main Maven repository used by
open-source developers to publish is Maven
Central.\footnote{\url{https://central.sonatype.com/}}

The study of dependency networks is well-established
\cite{decan2019empirical,fritz2024rustNetwork}, and several
datasets have been published to facilitate this over the
years~\cite{raemaekers2013maven,benelallam2019maven,
jaime2024goblin}. Goblin~\cite{jaime2024goblin} is the most
recent dataset, it is based on Maven Central and made available as a \textit{neo4j} graph
database~\cite{jaime2025challenge}. For our present study, we are using the 30-08-2024
version.\footnote{\href{https://zenodo.org/records/13683940}{goblin\_maven\_30\_08\_2024.dump}}

It is well-known that SECOs exhibit strong growth over time,
and this can be observed in the respective
networks~\cite{decan2019empirical,wittern2016look,
kikas2017structure, decan2016topology}. There are several
networks that can be considered here. Firstly, artifacts in
Maven are identified by a combination of group id (G), artifact
id (A) and version (V). In the context of network analysis, we
therefore refer to the versioned components as GAVs. Such GAVs
refer to other GAVs through the dependencies they declare. In
principle, dependencies can refer to sets of GAVs through the
use of version ranges, a feature intended to support semantic
versioning~\cite{semver}. However, this feature is rarely used
in Maven~\cite{dietrich2019dependency}, and it is difficult to accurately model
it as the resolution of dependency ranges hinges on the semantics of a particular build tool and the state of the repository at the time a component was built. 
We therefore decided to ignore dependencies declared to such ranges, and eliminated them from the dataset during dataset cleaning. We also removed GAVs for which we were unable to identify the release date. 
Of the 14,459,139 vertices and 119,660,406 edges in the dataset, 813,343 vertices (5.26\%) and 5,104,196 edges (4.26\%)  were ignored due to these issues. 

\begin{figure}[t]
\centering
\includegraphics[width=0.88\columnwidth]{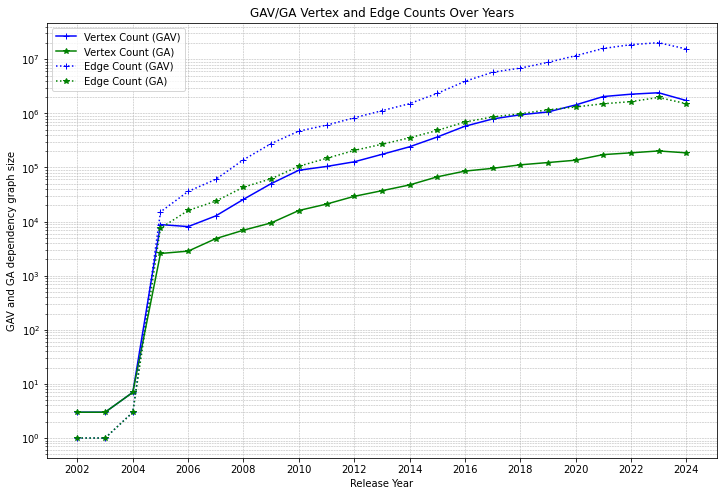}
\caption{Time evolution of vertex and edge counts for the GAV
and GA networks derived from Maven Central.}
\label{fig:counts}
\end{figure}

It is of interest to also consider the aggregation of GAVs into
unversioned components identified only by group and artifact
ids (GAs). Such GAs correspond to components, while GAVs
correspond to releases. A GA $ga_1$ depends on some other GA
$ga_2$ (i.e., there is a directed edge between $ga_1$ and
$ga_2$ in the GA graph), if some version $gav_1$ of $ga_1$
depends on some version $gav_2$ of $ga_2$. We consider a GA to
be released in a given year if any of its versions (GAVs) is
released in this year. Figure~\ref{fig:counts} summarizes the
growth of the GAV and GA networks over time. 

Organizations often have to make choices about which
programming languages to use for developing new products. One
of the deciding factors for that can be the availability of a
healthy ecosystem of open-source-software libraries. Desirable
attributes for such ecosystems include maturity, stability,
and positive evolution via continued maintenance and
innovation \cite{farshidi2021decision,dijkers2018exploring}.
Our study aims to provide insight into such fundamental
characteristics of SECOs. In particular, we focus on ways to
observe and quantify innovation as well as the dynamism of
popular artifacts.

Innovation is widely recognized as a key driver of growth \cite{mazzucato2015innovation, nicolaides2014research, ronkko2013innovation}. The Technical Committee of the International Organization for Standardization (ISO/TC 279) defines innovation as "a new or modified entity that creates or redistributes value." \cite{ISO56000:2025}. We use release types and frequencies to assess both maintenance (activity) and innovation. To gain a better understanding of the behavior of the key artifacts (elites) responsible for such innovation, we examine whether these elite artifacts remain static over the years as the ecosystem grows. In any given year, we define elite artifacts as those with more dependencies (usages) than their peers.

We use junit\footnote{\url{https://central.sonatype.com/artifact/junit/junit/versions}}and three other test frameworks—TestNG\footnote{\url{https://central.sonatype.com/artifact/org.testng/testng/versions}}, Mockito, and Spock\footnote{\url{https://central.sonatype.com/artifact/org.spockframework/spock-core/versions}}—as a case study. Adding tests is a widely used practice in open source development, and almost all projects have to choose testing libraries for this purpose. Among these, junit has been a fundamental component of software testing within the Maven ecosystem for many years, consistently ranking among the most influential Java projects.

In this paper, we study the observed growth of Maven Central in more detail. 
We start our investigation in Section~\ref{section:popularity} by
considering whether certain components dominate the ecosystem,
finding that this is generally the case. We then check whether
the set of those elite components is static or changes over time (Section~\ref{section:elites}). The observations from this
section motivate us to have a closer look at the innovation
dynamics of the Maven ecosystem, as explained in
Section~\ref{section:innovation}. A brief discussion of threats to validity (Section~\ref{section:threats}), related
work (Section~\ref{section:relatedwork}), and a conclusion
(Section~\ref{section:conclusion}) wrap up our contribution. 

\section{Popularity distribution}
\label{section:popularity}

A well-known effect in dynamic networks is the uneven
distribution of vertex degrees. This often reflects the uneven
spread of resources modelled by the network, such as
wealth in a population, links on the World Wide Web, citations
to scientific articles or patents etc.~\cite{albert2002network,
boccaletti2006network,vespignani2012network}.

For Maven, some components are extremely popular, such as
\textit{junit}, \textit{guava}, and various components from the
Apache commons family of components. On the other hand, there
are also many components that are not used by any other
component as a dependency. 

To characterize the popularity distribution and concentration of resources, we use
the \textit{Gini} index~\cite{dorfman1979gini}. Several studies
have employed this quantity to measure resource distributions in
software systems, including~\cite{decan2019empirical,
goloshchapova2013application,chelkowski2016inequalities,
wu2017gini}. The basic idea is to model incoming dependencies
as wealth. If few components attract most dependencies, this
suggests inequality and would result in a Gini value close
to $1$. 

In principle, such an analysis can be done on the level of either
the GAs or the GAVs. We find it more meaningful to consider the
GA network, as popular components tend to release new versions
more frequently as they are more actively maintained, i.e., when
versions (GAVs) are considered, incoming dependencies will be
split between those versions.  For instance, the popular
\textit{junit} framework has already released 11 versions between
January and October
2024.\footnote{\url{https://central.sonatype.com/artifact/org.junit.jupiter/junit-jupiter-api/versions}} Analysis of historical usage trends reveals that junit consistently ranked among the top 100 contributors to the ecosystem for an extensive period, spanning from 2005 to 2024, with an overall usage of 150,000 and counting, which far outpaces other testing frameworks, with the next closest project, TestNG, accumulating just over 11,900 usages. Here, we see junit's dominance when compared with its competitors.

To compute the Gini index for the GA network for a given year
$y_0$, we model population, wealth and ownership as follows:

\begin{enumerate}
\item \textbf{population:} the components (GAs) with any version
released in the year $y_0$ or before,
\item \textbf{wealth:} the  dependencies (edges) of components
released in the year $y_0$,
\item \textbf{ownership:} a component $ga_2$ owns a dependency
$gav_1 \rightarrow gav_2$ if there are versions $gav_2$ of
$ga_2$ and $gav_1$ of $ga_1$ such that $gav_1$ is released in
the year $y_0$.
\end{enumerate}

\begin{figure}[t]
\centering
\includegraphics[width=0.88\columnwidth]{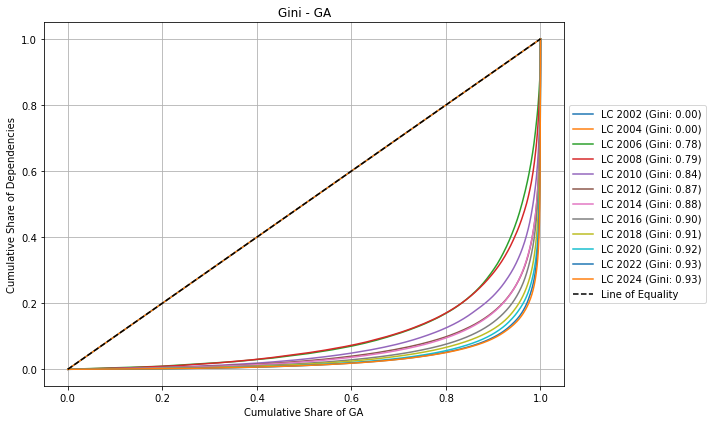}
\caption{\label{fig:gini} Lorenz curves (LC) and corresponding Gini
coefficients obtained for the dependencies released in a given
year and distributed among the existing GAs.}
\end{figure}

Figure~\ref{fig:gini} shows the respective Gini coefficients.
The measured values suggest significant inequality within the
distribution of dependencies, with the level of inequality
increasing over time. A superficial interpretation of this
trend could be that ``the rich get richer''. Expressed
differently, a few components could be dominating the network,
attracting more and more dependencies, and preventing new
components from becoming popular. We study next whether this
scenario is actually realized here. 

\section{Dynamism of elites}
\label{section:elites}

The observation of unevenness in the dependency distribution
suggests that the repository is dominated by a relatively small
number of components used by other components. To find out
whether the widely used (``elite'') components change over time,
we have studied whether and how the elite status of components
evolves in time. 

We define elite status on components (GAs) by ``wealth'' through in-degree as discussed in the last section. For each year, we are studying the top 10, top 100 and top 500 components. We then study how many components join and leave those elites in any given year.

There are some aspects here to consider, to ensure that
this analysis is sufficiently robust against confounding effects.
Components can be renamed (either the group id, or the
artifact id). To study this, we have used information of artifact
relocation from \textit{mvnrepository.com}. For example, for the
\textit{junit} artifact \textit{junit:junit}, this website contains an entry
stipulating that \textit{``this artifact was moved to: org.junit.jupiter:junit-jupiter-api''}.\footnote{\url{https://mvnrepository.com/artifact/junit/junit}}
We collected this information for all components (GAs) that had
elite status at some stage and created an alias map with such
renaming. This map has 4,000 entries. Using this map reduces
both the number of elite removals and additions only due
to renaming.

\begin{figure}
\centering
\includegraphics[width=0.88\columnwidth]{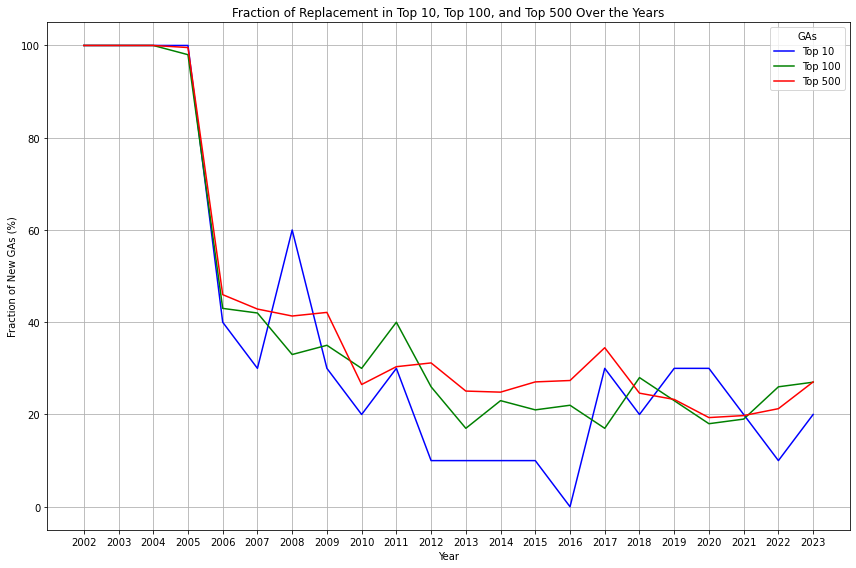}
\caption{Annual turnover in the composition of elite cohorts.}
\label{fig:replacement}
\end{figure}

Using these data, we looked at the changes in elites over the years. The results are depicted in Figure~\ref{fig:replacement}.  While the elites underwent some radical changes during the earlier years, this quickly stabilized around 2006. After some further gradual decline, we observe an annual turnover between 20 and 30\% since 2012. This suggests a mature ecosystem where popular components change at a stable rate over time.
For example, we see the increasing popularity of frameworks such as Mockito and Spock, which rank among the top 100 contributors between 2012 and 2024, just as TestNG became less popular and consequently fell out of the top 100 contributors in the year 2016. This further demonstrates the dynamic nature of the ecosystem as it has gained traction in recent years. In other words, this is an indication of ongoing innovation in Maven Central. We explore the notion of innovation further in the next Section. 

\section{Innovation}
\label{section:innovation}

To further study innovation~\cite{OECD2018innov}, we introduce two metrics that reflect complementary types of innovation. Those metrics are based on the following quantities measured for a given year: (1) \textbf{\textit{FirstGA}} is the number of components that had the first release of a version (GAV) during this year. (2) \textbf{\textit{LastGA}} is the number of GAs that have seen the last release in a given year. Measuring \textit{LastGA} for the last years of the time period studied is not meaningful, as there is still a reasonable chance that new versions will be released in the future. We therefore measure those values only up to 2022. Notably, once Maven components are added, they cannot be deleted.\footnote{This is a desirable property of a repository as withdrawing components from the repository can break dependencies and compromise downstream clients. For instance, this has led to the infamous \textit{leftpad} incident in npm~\cite{mens2016ecosystemic}.} (3) \textbf{\textit{MajorReleaseGA}} is the number of GAs that have seen a major version release in the given year. We assume here that major version releases contain new features and some innovation~\cite{Brown2024}, whereas minor and patch releases are mainly used for maintenance. 

\begin{figure}
\centering
\includegraphics[width=\columnwidth]{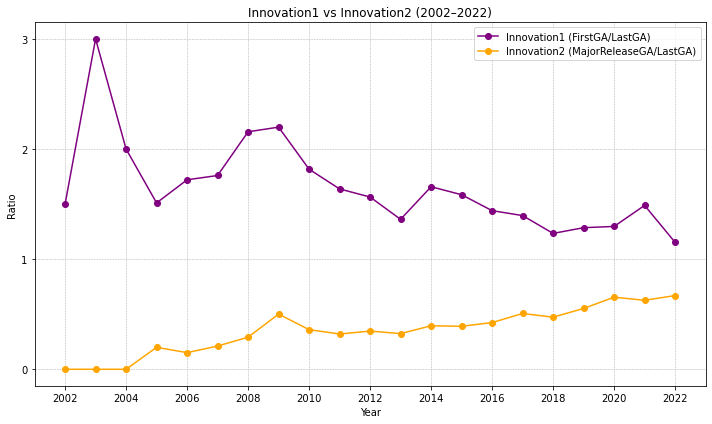}
\caption{Time evolution of two innovation metrics for Maven Central.}
\label{fig:innovation}
\end{figure}

The quantities \textit{FirstGA}, \textit{LastGA} and \textit{MajorReleaseGA} exhibit exponential growth over the considered time period, qualitatively similar to the growth pattern of the GA network displayed in Fig.~\ref{fig:counts}. We then consider 
$innovation1=FirstGA/LastGA$ and $innovation2=MajorReleaseGA/LastGA$
as two proxy measures for innovation. Normalization with respect to \textit{LastGA} is designed to measure innovative activity in the SECO relative to the noisy background of obsolete, or otherwise irrelevant, components present at any
given time. The results are depicted in Fig.~\ref{fig:innovation}. Interesting trends are seen to emerge in the time evolution of both $innovation1$ and $innovation2$ that are obscured by exponential growth when only the unnormalized quantities \textit{FirstGA} and \textit{MajorReleaseGA} are considered.

We observe a steady slow increase of $innovation2$ (innovation through major improvements of existing components) over time. Conversely, $innovation1$ (innovation through creating new components) is overall decreasing but appears to be stabilizing over time. Notably, $innovation1$ remains greater than one over the time period considered, which is an indication that still more components are being added to the ecosystem than components being abandoned. We interpret these trends as an indication that Maven Central is both healthy and mature as a SECO. 

\section{Related Work}
\label{section:relatedwork}

Decan et al. \cite{decan2019empirical} in their study on the empirical analysis of package dependency networks across seven packaging ecosystems (Cargo for Rust, CPAN for Perl, CRAN for R, npm for JavaScript, NuGet for the .NET platform, Packagist for PHP, and RubyGems for Ruby) identified common challenges in these ecosystems. They accessed the growth, changeability, reusability, and fragility of these ecosystems, revealing trends of network expansion, the centrality of a small number of packages in driving updates, and the prevalence of fragile packages with numerous transitive dependencies. Similarly to our approach, their study performed a dynamic analysis using the Gini index as a key metric to explore inequality and concentration within these networks. However, while their study spans a diverse range of ecosystems, differing in size, age, and policies, our present research is more focused, dealing solely with Maven's ecosystem.
Other researchers often use popularity metrics to sample datasets or investigate software properties. While some studies have described the popularity of software components in terms of social characteristics, others have described them in terms of technical aspects~\cite{Zerouali2014}. For instance, a study of GitHub developers conducted by Lee et al. \cite{lee2013github} demonstrated that very well-known developers, who are often referred to as "rock stars", have a greater influence on the projects their followers contribute to. In comparison to our study which explores the network evolution and growth in the Maven ecosystem, they conducted a dynamic analysis of how the actions and interactions of developers evolve.
Borges et al. \cite{borges2018s} conducted a survey involving 400 Stack Overflow users. The results from their poll showed that the users viewed GitHub metrics such as stars, forks, and watchers as highly valuable indicators of how popular a project is. Furthermore, the majority of the comments from OSS developers questioned by Bogart et al. \cite{bogart2016break} on why they chose the right dependencies for their software projects fell into groups pertaining, reputation and popularity in the community. 

Other researchers like Sajnani et al. \cite{sajnani2014popularity} who measured the popularity of 2,406 Maven components by analyzing how often they were used in 55,191 open-source Java projects, have also argued that usage of software components can be used as a measure of their popularity. Their interpretations work under the assumption that if a component is widely (re)used, then it is generally regarded as good.

The shortcomings of using social attributes of software components as a popularity metric are exacerbated by the fact that they cannot provide a complete picture of real usage, as they can be easily influenced by individual's preferences or trend \cite{papamichail2019measuring}. Research conducted in the past by Kitchenham et al. \cite{kitchenham1988evaluation}, Fenton et al. \cite{fenton1999critique}, and Vasa et al. \cite{vasa2007inevitable} has demonstrated how widely skewed software metrics are in general, making accurate interpretation with conventional descriptive statistical analysis challenging.

\section{Threats to Validity and Reproducibility}
\label{section:threats}

\subsection{Threats to Validity}

There might be some additional patterns that could influence the accuracy of the analysis of elites in Section~\ref{section:elites}. Notably, some components split into smaller components. We currently consider those modules as proper new components, but one could argue that some modules "inherit" the component status from their respective parents, and should be treated like aliases resulting from relocating components. 

We decided not to model version ranges. The impact of this decision is small, as discussed in Section~\ref{section:introduction}.

In Section~\ref{section:innovation}, we have made the assumption that major releases correspond to innovation, i.e., generally entail the introduction of new features. This is consistent with the objectives of semantic versioning~\cite{semver}. We note that many projects do not strictly follow semantic versioning~\cite{raemaekers2017semantic,ochoa2022breaking}. However, this mainly relates to the presence of breaking changes in non-major releases. 

\subsection{Reproducibility of Results}

The scripts and datasets created and used in this study are available on GitHub.\footnote{ \url{https://github.com/nkiru-ede/Popularity_and_Innovation_in_Maven_Central/releases/tag/MSR25v1.0}}

\section{Ethical Implications}
\label{section:ethics}

In attempts to measure lofty concepts such as utility,
popularity, innovation and the like, researchers typically
introduce noisy proxy measures of such socially constructed
abstractions. One needs to reflect on limitations and inherent
biases of the adopted quantities before attempting to interpret
these more widely and derive deeper meaning, or even policy
directions, from their cross-correlation. Furthermore,
characterizing random distributions by only a few summary
quantities (e.g., the Gini index) can lead to misrepresentations
of diversity and variability in human endeavor. Quantitative
analyses need to be augmented with qualitative insights from
representative practitioners (in our case, software developers)
to establish the real driving forces behind the trends exhibited
in the data.

\section{Conclusion}
\label{section:conclusion}

We have studied the distribution of dependencies in Maven Central over 22 years (2002-24), using the \textit{Goblin} dataset. We find that Maven Central is dominated by a relatively small number of components that attract most dependencies. This in itself is hardly surprising. However, interestingly, the set of elite components is highly dynamic and exhibits significant annual turnover. We also observe that there is a stable rate of renewal through innovation in Maven Central.


\end{document}